\documentclass[aps,epsf,reprint,twocolumn]{revtex4-2}
\usepackage{graphicx}
\usepackage{dcolumn}
\usepackage{amsmath}
\usepackage{amssymb}
\usepackage{empheq}
\usepackage[colorlinks]{hyperref}
\hypersetup{citecolor=blue}

\usepackage{bm}
\usepackage{epstopdf}
\usepackage{amsthm}
\usepackage{float}
\graphicspath{{./figures/}}
\usepackage{accents}
\usepackage{comment}

\def\epsilon{\varepsilon}

\def\epsilon{\varepsilon}

\begin{document}


\title{Traveling vegetation-herbivore waves can sustain ecosystems threatened by droughts and population growth}

\author{Joydeep Singha}
\email{joydeep@post.bgu.ac.il}
\affiliation{The Swiss Institute for Dryland Environmental and Energy Research, Blaustein Institutes for Desert Research, Ben-Gurion University, Midreshet Ben-Gurion 8499000, Israel}
\author{Hannes Uecker}
\affiliation{Institute for Mathematics, Carl von Ossietzky University Oldenburg, P.F 2503, 26111 Oldenburg, Germany}
\author{Ehud Meron}
\affiliation{The Swiss Institute for Dryland Environmental and Energy Research, Blaustein Institutes for Desert Research, Ben-Gurion University, Midreshet Ben-Gurion 8499000, Israel}
\affiliation{Physics Department, Ben-Gurion University, Beer Sheva 8410501, Israel}
\date{\today}

\begin{abstract}
Dryland vegetation can survive water stress by forming spatial patterns but is often subjected to herbivory as an additional stress that puts it at risk of desertification. Understanding the mutual relationships between vegetation patterning and herbivory is crucial for securing food production in drylands, which constitute the majority of rangelands worldwide. Here, we introduce a novel vegetation-herbivore model that captures pattern-forming feedbacks associated with water and herbivory stress and a behavioral aspect of herbivores representing an exploitation strategy.
Applying numerical continuation methods, we analyze the bifurcation structure of uniform and patterned vegetation-herbivore solutions, and use direct numerical simulations to study various forms of collective herbivore dynamics. 
We find that herbivory stress can induce traveling vegetation-herbivore waves and uncover the ecological mechanism that drives their formation. 
In the traveling-wave state, the herbivore distribution is asymmetric with higher density on one side of each vegetation patch. At low precipitation values their distribution is localized, while at high precipitation the herbivores are spread over the entire landscape.
Importantly, their asymmetric distribution results in uneven herbivory stress, strong on one side of each vegetation patch and weak on the opposing side -- weaker than the stress exerted in spatially uniform herbivore distribution. Consequently, the formation of traveling waves results in increased sustainability to herbivory stress.
We conclude that vegetation-herbivore traveling waves may play an essential role in sustaining herbivore populations under conditions of combined water and herbivory stress, thereby contributing to food security in endangered regions threatened by droughts and population growth. 
\end{abstract}

\maketitle

\section{\label{intro}Introduction}
Understanding the response of ecosystems to climate change is crucial for maintaining essential provisioning services that ecosystems provide to humans. Of these, feeding livestock by grazing in dryland pastures stands out in its scope; drylands constitute 78\%  of rangelands worldwide and are home to about billion people who rely on grazing as a critical source of protein and income~\cite{Maestre2022science}. Grazing, in turn, may have complex positive and negative feedback effects on dryland vegetation in terms of carbon stocks~\cite{Munjonji2020frontiers,Deng2023carbon_res}, species diversity and ecosystem functioning~\cite{Hanke2014ecol_appl,Nathan2016j_ecology,Anvar2023functional_ecology,Gaitan2018ldd}. 

One of the hallmarks of dryland landscapes is their patchy nature, often appearing as mosaics of vegetation and ``bare-soil'' patches, devoid of vegetation but not of other life forms. This is a pattern formation phenomenon driven by positive feedback loops between vegetation growth and water transport~\cite{Meron2018ann_rev,Meron2019pt}. In homogeneous areas, these patterns can be strikingly ordered, in line with the predictions of pattern formation theory~\cite{Cross1993rmp,Meron2015book}. Well-reported examples are the so-called ``tiger bush'' patterns, consisting of parallel stripes of woody vegetation on sloped terrains oriented perpendicular to the slope direction~\cite{Valentin1999catena,Lefever1997bmb,Deblauwe2012eco_mono,Bastiaansen2018pnas}, and ``fairy circles'', which are hexagonal patterns of bare-soil gaps in grasslands, observed in flat terrains~\cite{Cramer2013plosone,Tschinkel2015plosone,getzin2016pnas,Tarnita2017nature,Guirado2023pnas,Bennett2023pnas}. Vegetation pattern formation is a mechanism by which plant populations or communities tolerate water stress, as vegetation patches benefit from an additional water resource -- the unutilized rain that falls in adjacent bare-soil patches~\cite{Meron2019pt}. This mechanism is likely to apply to non-dryland ecosystems as well, as many of them are experiencing a drying trend~\cite{Grunzweig2022nat_eco_evo}.

Herbivores exert additional stress on already water-stressed dryland pastures, putting these ecosystems at risk of desertification and loss of function~\cite{Maestre2022science}. However, despite this concern and the high interest in understanding plant-herbivore interactions~\cite{Wetzel2023ann_rev}, very little is known about the mutual relationships between vegetation patterning and behavioral aspects of herbivore dynamics. Basic questions, such as how herbivore grazing or browsing affects vegetation patterns or how foraging strategies in patterned-vegetation landscapes affect 
herbivore survival, have hardly been addressed. 

Mathematical models have been highly instrumental in understanding the mechanisms that drive vegetation pattern formation~\cite{Klausmeier1999science, Okayasu2001,HilleRisLambers2001ecology,Hardenberg2001prl,Lejeune2002pre,Rietkerk2002an,Gilad2004prl,Borgogno2009gr} and the roles these patterns play in maintaining ecosystem resilience to droughts~\cite{Siteur2014eco_comp,Zelnik2021plos_comp_biol,Rietkerk2021science} and biodiversity~\cite{Bera2021elife,Guill2021eco_let,Bennett2023pnas_nexus}. 
The importance of model studies stems in part from the difficulty of conducting controlled laboratory experiments on dynamical behaviors, such as instabilities and state transitions, because of the long time and space scales associated with vegetation patterning. 

Mathematical models can be equally instrumental in unraveling the complex relationships between vegetation patterning and herbivore dynamics, and their implications for biodiversity and ecosystem function. To this end, the models 
should capture water-vegetation feedbacks capable of inducing vegetation patterns, and spatially explicit herbivore dynamics that include behavioral elements, such as foraging strategy~\cite{Owen-Smith2010ptrsb}. A few spatially-explicit vegetation-herbivore models have been proposed and studied recently, but none of them meets both requirements~\cite{Siero2019amnat,Fernandez-Oto2019eco_comp, Ge2022eco_let, Pal2023pre}. 

In this paper, we introduce a novel vegetation-herbivore model (Section \ref{sec:model})
that captures pattern-forming water-vegetation feedback and a behavioral element -- preferential herbivore movement to nearby areas of denser vegetation~\cite{Osem2004j_ecology}, reflecting an exploitation strategy~\cite{Berger-Tal2014plosone}. We refer to this behavior as ``vegetaxis'', in analogy to chemotaxis -- the movement of micro-organisms up gradients of chemical stimuli~\cite{Keller1971jtb,Wadhams2004nature}, and prey-taxis~\cite{Kareiva1987amnat}. 
Using this model, we study the impacts of herbivore dynamics on vegetation patterning (Section \ref{sec:traveling_waves}), and the reciprocal effects of vegetation patterning on herbivore survival in water-stressed landscapes  (Section \ref{sec:survival}). We conclude with a discussion of the implications of our results to plant-species diversity and ecosystem functioning (Section \ref{sec:Discussion}).


\section{A spatially explicit model for vegetation-water-herbivore dynamics}
\label{sec:model}
Several scale-dependent feedbacks capable of inducing vegetation patterns in water-limited flat terrains have been proposed~\cite{Meron2019pt}. They differ by the dominant form of water transport through which plants draw water from their surroundings: overland water flow, water conduction by laterally spread roots, and lateral soil-water diffusion~\cite{Meron2018ann_rev,Bennett2023pnas_nexus}. For simplicity, we consider here the latter water-transport form -- lateral soil-water diffusion -- which applies to ecosystems with sandy soil and plant species with laterally confined root zones. This choice allows us to reduce the water-vegetation equations to a system of two local partial-differential equations for the vegetation above-ground biomass density $B(\mathbf{X}, T)$[kg/m$^2$] and the soil-water content $W(\mathbf{X}, T)$[kg/m$^2$]~\cite{Zelnik2015pnas}, where $\mathbf{X} = (X, Y)$ represents the spatial coordinates and $T$ is the time coordinate. The main results reported here are not expected to depend on this particular choice of water transport form. 
Complementing the equations for the vegetation biomass and water variables is an equation for the herbivore biomass density, $H(\mathbf{X}, T)$[kg/m$^2$]. The model then reads:
\begin{align}
    \begin{split}
        \partial_TB&=\Lambda BW(1+EB)^2(1-B/K_B)-M_BB+D_B\nabla^2B\\
        &-G(B)H,\\
        \partial_TW&=P-\frac{NW}{1+RB/K_B}-\Gamma BW(1+EB)^2 + D_W\nabla^2W,\\
        \partial_TH&=-M_HH+A G(B)H{\left(1-H/K_H\right)} -\nabla\cdot J_H,\\
    \end{split}
    \label{eq:BWH}
\end{align}
where $\nabla=\hat{\textbf{x}}\partial_X+\hat{\textbf{y}}\partial_Y$ and $\hat{\textbf{x}}, \hat{\textbf{y}}$ are unit vectors in the $X,Y$ directions, respectively. We refer the reader to Table 1 for a brief description of all model parameters, and their values unless otherwise stated. Plant biomass ($B$) grows due to water-dependent reproduction, enhanced by root growth ($\Lambda W(1+EB)^2$) and attenuated by limiting factors such as pathogens ($1-B/K_B$). Plant biomass decreases due to natural mortality ($-MB$) and herbivory ($-G(B)H$), and is distributed in space by short-range seed dispersal ($D_B\nabla^2B$). Soil water content increases due to precipitation ($P$), and decreases due to biomass-dependent evaporation, which captures the effect of shading ($NW/(1+RB/K_B)$) and water uptake by plants' roots ($-\Gamma BW(1+EB)^2$). Soil-water content is distributed in space by lateral diffusion ($D_W\nabla^2W$). 

The new elements in the model are related to the herbivore dynamics. Herbivore biomass ($H$) grows due to biomass-dependent reproduction ($A G(B)H(1-H/K_H)$) and decreases due to natural mortality ($-M_HH$). The herbivore per-capita grazing or browsing rate, $G(B)$, is biomass-dependent to account for biomass-limited herbivory at low plant-biomass densities and satiation at higher densities and is given by
\begin{equation}
G(B)= \frac{\alpha B}{\beta + B}\,,
\label{eq:G}
\end{equation}
where $\alpha$ is the maximal consumption rate of vegetation biomass per unit herbivore density and $\beta$ is the characteristic vegetation biomass for herbivore satiation. Smaller values of $M_H,~\alpha$ and $\beta$ account for circumstances where herbivore feeding is increasingly dependent on fodder as a supplementary food resource.
As two herbivores cannot occupy the same location, there is a limit to their density, quantified by  $K_H$. 

The spatial distribution of herbivores is described by the flux 
\begin{equation}
\label{eq:J_H}
    J_H = -D_R(B)\nabla H + HD_V(B)\nabla B\,.
\end{equation}
The first term describes random walk with a motility rate that depends on the vegetation biomass and is given by 
\begin{equation}
    D_R(B) = D_{HH}\frac{\xi^2}{\xi^2 + B^2}\,,
    \label{eq:D_R}
\end{equation} 
where $D_{HH}$ and $\xi$ are constants (see Table 1). This form accounts for the slowing down of herbivore movement in areas where herbivores begin to sense vegetation. The second term in (\ref{eq:J_H}) describes vegetaxis, that is, herbivore motion up gradients of vegetation biomass, with a motility rate
\begin{equation}
    D_V(B) = D_{HB}\frac{\kappa}{\kappa + B}\,.
    \label{eq:D_V}
\end{equation}
The consideration of vegetaxis is motivated by the finding that the size and height of plants are good predictors of species’ sensitivity to grazing~\cite{Noy-Meir1989j_ecology,Osem2004j_ecology,Nathan2016j_ecology}. 
The biomass dependence of the motility $D_V(B)$ reflects the assumption that the herbivores slow down their search once they sense tails of vegetation-biomass distributions: $B\approx \kappa$.

We study the model equations (\ref{eq:BWH}) using numerical continuation methods in one space dimensions (1D) and direct numerical simulations in 1D and 2D. We use the continuation software MatCont~\cite{matcont} for ordinary differential equations to obtain the bifurcation diagrams in Fig. \ref{fig:BODE_bif} of spatially uniform solutions, and the continuation software pde2path~\cite{p2pbook2021} for partial differential equations to obtain the bifurcation diagrams in Fig. \ref{fig:bif_diag_alpha}, \ref{fig:bif_diag_alpha_high}, \ref{fig:bif_diag_M_H}, \ref{fig:TW_merging} and \ref{fig:P_DHB}b. 
The stability of solution branches was calculated by extracting the solutions and applying numerical stability analysis. 
The phase diagrams in Fig. \ref{fig:P_alpha} and \ref{fig:P_DHB}a were computed using numerical fold-,branch- and Hopf-point continuation in pde2path. Direct numerical simulations to produce 
Fig. \ref{fig:TW_spacetime} and \ref{fig:2d_simulations} were carried out using two Python packages for solving partial differential equations, py-pde \cite{Zwicker2020} and Dedalus \cite{dedalus2020}. 
 Table 1 displays the parameter values used in these numerical studies. Since our goal in this study is to highlight general behaviors, such as bifurcation structures, shared by many different systems, we do not attempt in this choice of parameter values to model any particular system.

\begin{figure}[ht]
    \centering
    \includegraphics[width=0.35\textwidth]{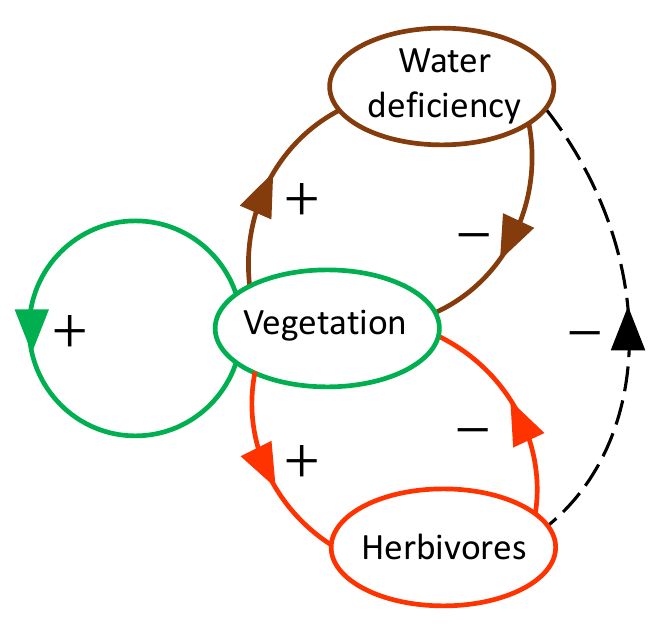}
    \caption{\footnotesize An illustration of the model as an activator-two-inhibitors system. Vegetation (activator), enhances its own growth via reproduction (green loop), creates water scarcity via water uptake (positive brown feedback), and enhances herbivore growth by providing food (positive orange feedback). Water scarcity (1$^{st}$ inhibitor) and herbivores (2$^{nd}$ inhibitor) inhibit vegetation growth (negative brown and orange feedbacks); the former by water uptake and the latter by herbivory. The negative feedback of herbivores on vegetation, combined with the positive feedback of vegetation on water scarcity result in a negative feedback of herbivores on water scarcity (dashed black line).  } 
    \label{fig:AII}
\end{figure}

The model equations (\ref{eq:BWH}) represent an activator-two-inhibitors system. The activator is the vegetation biomass and the two inhibitors are water scarcity and herbivores. As the solid lines in Fig. \ref{fig:AII} illustrate, vegetation growth is an autocatalytic process that enhances itself by reproduction, drives soil water depletion by water uptake, and herbivore-population growth by herbivory. Depleted water and herbivory, in turn, inhibit vegetation growth. The dashed line in Fig. \ref{fig:AII} represents the indirect effect herbivores have on soil-water content; by reducing vegetation biomass (leaf area) herbivores decrease soil-water uptake and thereby increase soil-water content (decrease soil-water depletion).

At low precipitation values, soil-water content is expected to be the dominant inhibitor of vegetation growth, while at high precipitation values, herbivores are expected to be the dominant inhibitor.  Since both inhibitors diffuse much faster than seed dispersal, scale-dependent feedbacks capable of forming spatial patterns, may emerge. 

A dimensional analysis leading to a non-dimensional form of the model equations (\ref{eq:BWH}) is described in Appendix A. 

\bigskip
\begin{widetext}
\begin{tabular}{|p{1cm}|p{11.5cm}|p{4cm}|}
\hline
\multicolumn{3}{|c|}{Table 1. List of the parameters, their description, their values, and their units.} \\
\hline
$P$ &Precipitation rate &variable,~$\mathrm{kg}\cdot \mathrm{m}^{-2}\cdot \mathrm{y}^{-1}$,\\ & & or equivalently $\mathrm{ mm\cdot y^{-1}}$\\
$\Lambda$ &Growth rate per unit soil water &$\mathrm{ 0.5~m^2\cdot kg^{-1}\cdot y^{-1}}$\\
$E$&Root to shoot ratio&$\mathrm{10~m^2\cdot kg^{-1}}$\\
$K_B$&Maximal standing vegetation biomass&$\mathrm{0.9~kg\cdot m^{-2}}$\\
$M_B$&Plant mortality rate in dry soil in the absence of herbivores&$\mathrm{11.4~y^{-1}}$\\
$D_B$&Seed dispersal rate&$\mathrm{1.2~m^2\cdot y^{-1}}$\\
$N$&Evaporation rate&$\mathrm{20~y^{-1}}$\\
$R$&Reduced evaporation due to shading&0.01\\
$\Gamma$&Soil-water consumption rate per unit biomass & 10~$\mathrm{m^2\cdot kg^{-1}\cdot y^{-1}}$\\
$D_W$&Lateral soil water diffusion coefficient&$\mathrm{150~m^2\cdot y^{-1}}$\\
$M_H$&Herbivore mortality rate in the absence of herbivory&$\mathrm{0.06~y^{-1}}$\\
$A$&Fraction of consumed vegetation used for herbivore growth&0.3\\
$\alpha$&Maximal vegetation-consumption rate per unit herbivore density&0.6~$\mathrm{y}^{-1}$\\
$\beta$&
Satiation biomass&$\mathrm{0.82~kg\cdot m^{-2}}$\\
$K_H$&Maximal herbivore biomass&$\mathrm{175~kg\cdot m^{-2}}$\\
$D_{HH}$&Maximal random motility &$\mathrm{400~m^2\cdot y^{-1}}$\\
$\xi$&Reference vegetation biomass at which random motility  drops by $50\%$ & $\mathrm{10^{-3}~kg\cdot m^{-2}}$\\
$D_{HB}$&Maximal vegetaxis motility&$\mathrm{700~m^2\cdot kg^{-1}\cdot y^{-1}}$\\
$\kappa$&Reference vegetation biomass at which vegetaxis motility drops by $50\%$&$\mathrm{10^{-4}~kg\cdot m^{-2}}$\\
\hline
\bigskip
\end{tabular}
\end{widetext}

\section{Herbivores can induce traveling vegetation waves}
\label{sec:traveling_waves}
According to vegetation pattern formation theory, vegetation patterns in flat terrains and constant environments are stationary, arising from a nonuniform (Turing) instability of uniform vegetation as the precipitation $P$ drops below a critical value $P_T$~\cite{Rietkerk2008tree,Meron2019pt}. In this section, we show that herbivores can induce traveling vegetation-herbivore waves in two distinct ways. At low herbivory stress, the traveling waves emerge from an oscillatory instability of stationary vegetation patterns, with herbivores localized at one side of each vegetation patch. At sufficiently higher herbivory stress, the traveling waves emerge from a non-uniform oscillatory instability of uniform vegetation-herbivore state. In this case, the herbivores are non-uniformly spread all over the system, attaining maximal density values at one side of each vegetation patch. We further explain the ecological mechanism by which traveling vegetation-herbivore waves appear and demonstrate richer traveling-wave behaviors that the model predicts. We begin with analyzing spatially uniform solutions from which stationary and traveling patterns emerge.

\subsection{\label{sec_1}Spatially uniform model solutions}
The model (\ref{eq:BWH}) has three stationary uniform solutions $(B,W,H)$ describing bare soil $BS:(0,W_{BS},0)$, uniform vegetation without herbivores $UV:(B_{UV},W_{UV},0)$, and uniform vegetation with herbivores $UH:(B_{UH},W_{UH},H_{UH})$. The $BS$ solution is given by $W_{BS}=P/N$. Analytic expressions for the $UV$ and $UH$ solutions are lengthy as they involve solving a quartic equation for $B$ and are not displayed here. We, therefore, resort to numerical bifurcation analysis.

\begin{figure}[ht]
    \centering
    \includegraphics[width=0.45\textwidth]{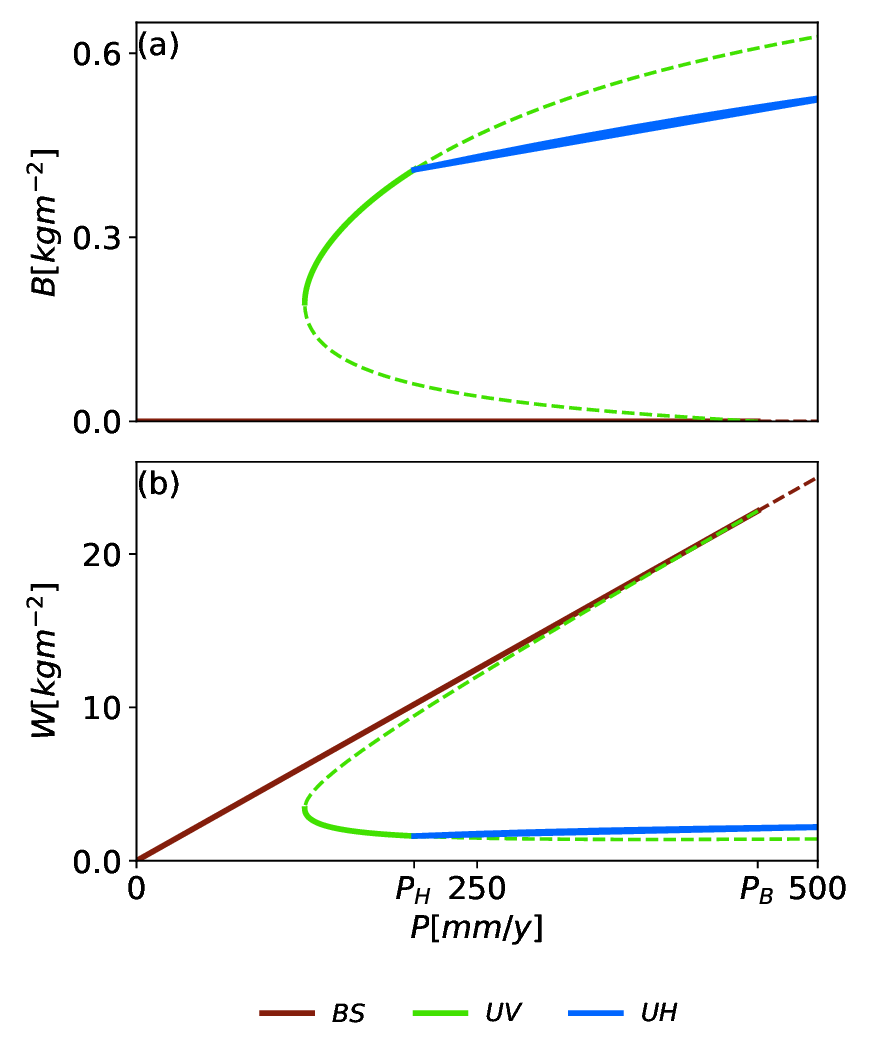}
    \caption{\footnotesize Bifurcation diagrams of stationary uniform solutions of Eq. (\ref{eq:BWH}) along the rainfall gradient. (a) A diagram showing the vegetation biomass $B$, (b)  a diagram showing the soil-water content $W$. The labels $BS,~UH,~UV$ denote, respectively, the bare soil solution and uniform vegetation solutions with and without herbivores. Solid (dashed) lines denote stable (unstable) solutions.  The label $P_B$ denotes the precipitation threshold where the $BS$ solution bifurcates to the $UV$ solution, while the label $P_H$ denotes the precipitation threshold where the $UV$ solution bifurcates to the $UH$. Parameters: as in Table 1.
    }
    \label{fig:BODE_bif}
\end{figure}

\begin{figure*}[t!]
    \centering
    \includegraphics[width=0.436\textwidth]{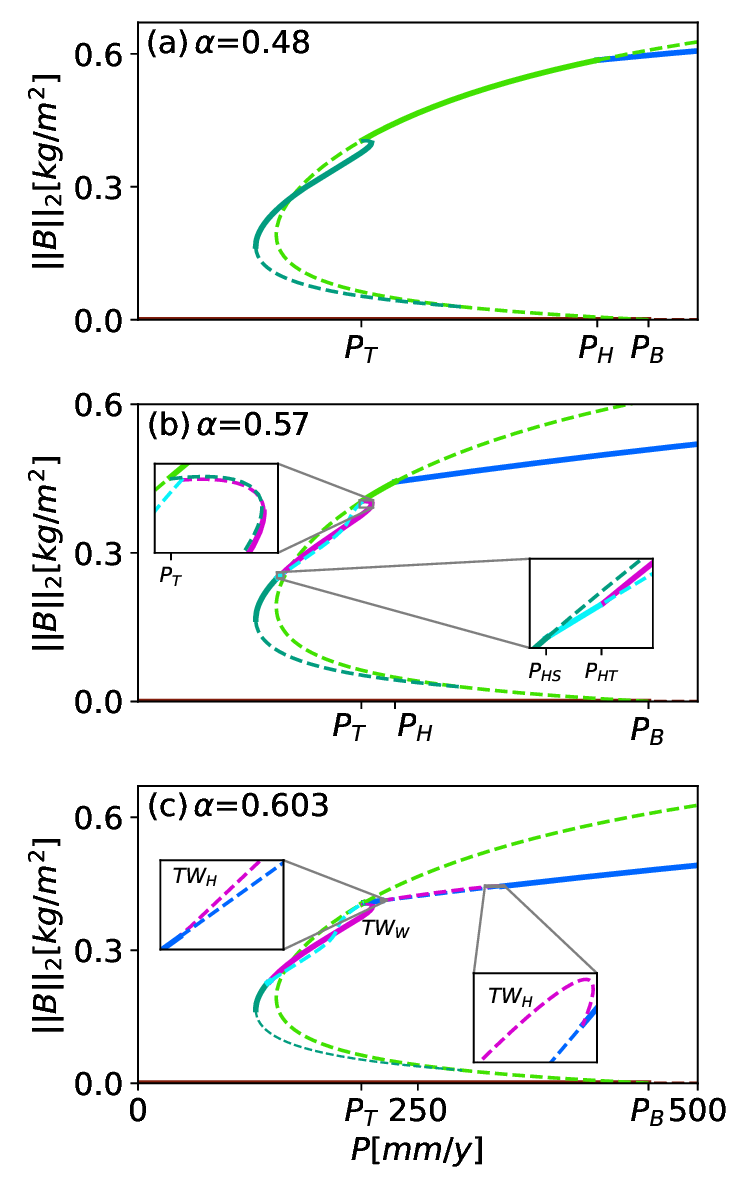}
    \includegraphics[width=0.23\textwidth]{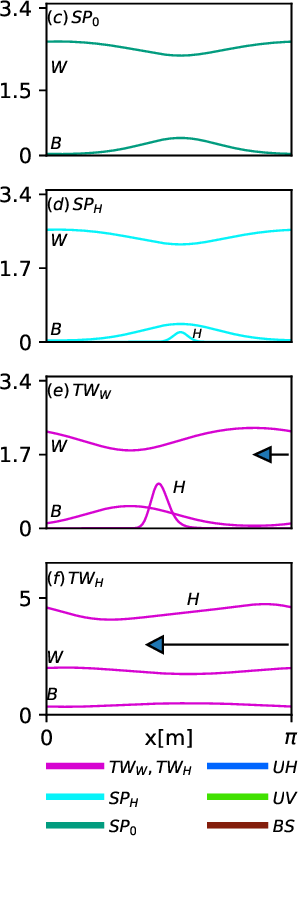}
    \caption{\footnotesize Bifurcation diagrams of uniform and 1D nonuniform solutions of Eq. \eqref{eq:BWH} for low herbivory stress. (a) At sufficiently low maximal-consumption rates $\alpha$, the uniform vegetation solution ($UV$, light green) loses stability to the uniform vegetation-herbivore state ($UH$, blue) at $P=P_H$ and to stationary periodic vegetation patterns devoid of herbivores ($SP_0$, dark green) at $P=P_T<P_H$. (b) At higher $\alpha$, the stationary vegetation patterns $SP_0$ lose stability to stationary vegetation patterns with low herbivore density ($SP_H$, light blue) as $P$ exceeds $P_{HS}$, which at a slightly higher threshold, $P_{HT}$, lose stability to traveling vegetation-herbivore waves ($TW_W$, magenta). (c) At yet higher $\alpha$, the uniform vegetation-herbivore state ($UH$, blue) loses stability to a different traveling vegetation-herbivore wave solution ($TW_H$, magenta). This solution is unstable when it appears but becomes stable at higher $P$ values (see Fig. \ref{fig:bif_diag_alpha_high}).
    The vertical axis represents the L2 norm of the biomass. Solid (dashed) lines denote stable (unstable) solutions. 
    The panels on the right show examples of the four spatially periodic solutions that appear in panel c, over one period: (c) $SP_0$ ($P = 114$ mm/y), (d) $SP_H$ ($P = 116$ mm/y), (e) $TW_W$ ($P = 153.2$ mm/y), and (f) $TW_H$ ($P = 262.24$ mm/y). The black arrows in panels e,f denote the traveling wave direction. Their different length reflects the higher speed of the $TW_H$ solution compared to the $TW_W$ solution. Parameters are as indicated and as in Table 1.
    }
    \label{fig:bif_diag_alpha}
\end{figure*}

\begin{figure*}[t!]
    \centering
    \includegraphics[width = 0.8\textwidth]{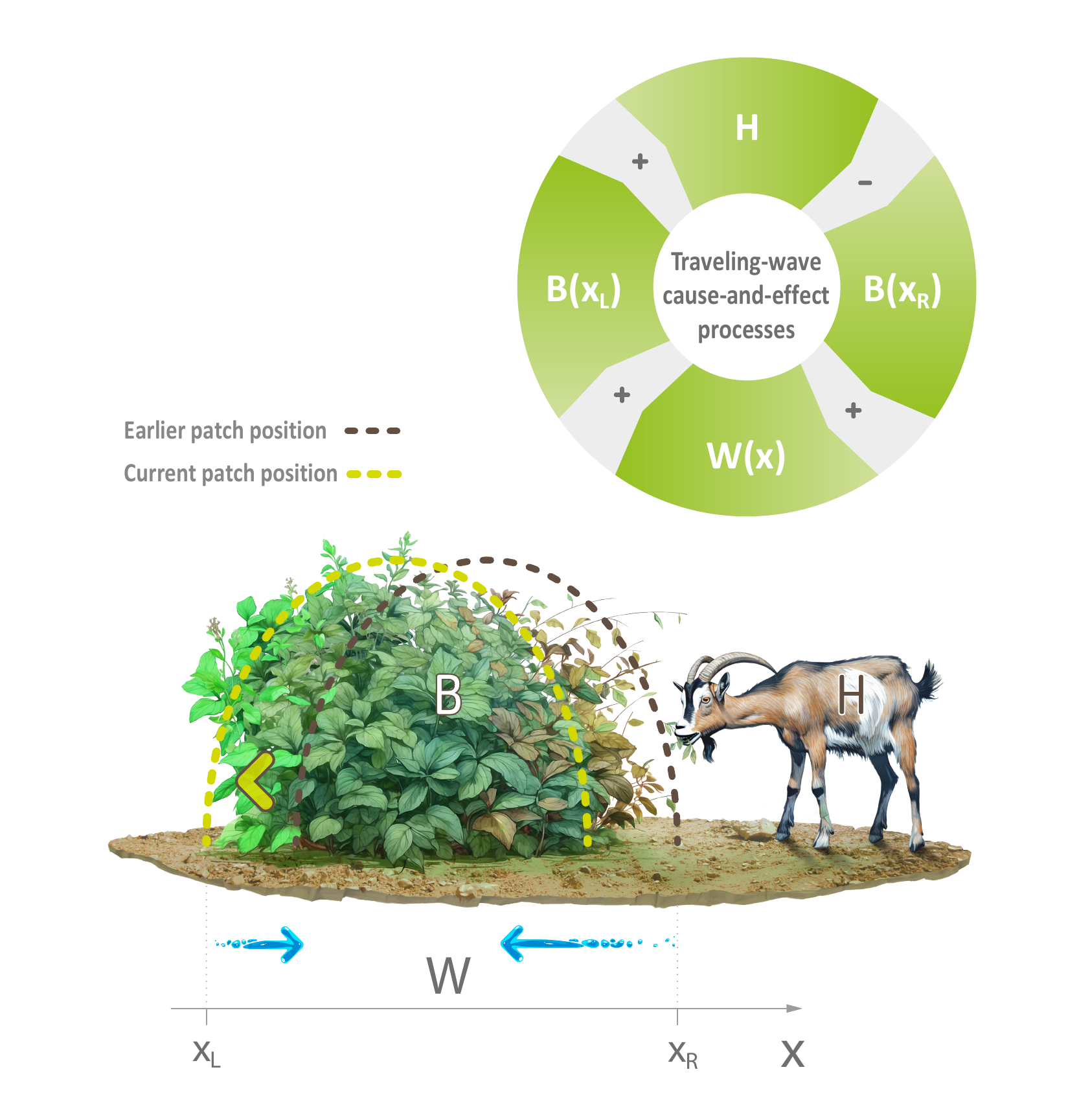}
    \caption{\footnotesize A schematic illustration of the mechanism that drives the formation of traveling vegetation-herbivore waves. Herbivory on the right side ($X_R$) of a vegetation patch induces a series of negative (-) and positive (+) feedbacks, as the loop of cause-and-effect processes at the top right part of the figure illustrates, resulting in a displacement of the patch to the left. The blue arrows and their lengths denote increased soil-water flux from the right patch side to the patch center and decreased flux from the left side to the center. A detailed explanation appears in the text.
    }
    \label{fig:TW_mechanism}
\end{figure*}

Figure \ref{fig:BODE_bif} shows bifurcation diagrams of the stationary uniform solutions $BS,~UV,~UH$ where the bifurcation parameter is the precipitation rate $P$. Figure \ref{fig:BODE_bif}a shows the biomass component of the solutions whereas Fig. \ref{fig:BODE_bif}b shows the soil-water component. The diagrams show the existence and stability ranges of the solutions, uncovering the following sequence of stable states as precipitation increases: $BS$ at precipitation rates too low to sustain vegetation, $UV$ at precipitation rates high enough to sustain vegetation but not sufficiently high to sustain herbivores and $UH$ at precipitation rates high enough to sustain both vegetation and herbivores. They further reveal a bistability range of $BS$ and $UV$ at low precipitation and a bistability range of $BS$ and $UH$ at higher precipitation. The latter range terminates at the bare-soil instability threshold $P=P_B=NM/\Lambda$. Beyond this threshold $UH$ is the only stable uniform state. 
The soil-water diagram (Fig. \ref{fig:BODE_bif}b) reveals a significant aspect of the $UH$ solution; the presence of herbivores increases the soil water content ($W_{UH}>W_{UV}$). This is because of the lower water uptake rate of grazed vegetation, which has lower biomass density, as Fig. \ref{fig:BODE_bif}a shows ($B_{UH}<B_{UV}$). In nature, this is a result of reduced transpiration due to a smaller total leaf area~\cite{Golluscio2022jae}.

\subsection{The onset of periodic traveling vegetation-herbivores waves}
In the absence of herbivores, the scale-dependent water-vegetation feedback that the model captures can induce a Turing instability of uniform vegetation to stationary periodic patterns, as the precipitation rate drops below a threshold value~\cite{Rietkerk2008tree,Meron2012ecol_mod}. The question we address here is whether and how the presence of herbivores affects vegetation patterning given the assumptions we have made on herbivore movement as formulated mathematically by the herbivore flux $J_H$ (Eq. (\ref{eq:J_H}-\ref{eq:D_V})). That is, herbivores move fast and randomly in bare-soil or sparsely-vegetated areas, move preferentially to areas of denser vegetation, and slow down as the vegetation becomes denser~\cite{Noy-Meir1989j_ecology,Osem2004j_ecology}. Other factors that may affect herbivore movement are discussed in section \ref{sec:Discussion}.  We address this question by considering increasing levels of herbivory stress.

Herbivores stress in the model can be quantified by the maximal vegetation-consumption rate $\alpha$ (see Eq. \eqref{eq:G}). Figure \ref{fig:bif_diag_alpha} shows bifurcation diagrams of spatially uniform and nonuniform solutions of Eq. \eqref{eq:BWH} in one space dimension (1d), with the precipitation rate $P$ as a bifurcation parameter.
At sufficiently low $\alpha$ values (Fig.  \ref{fig:bif_diag_alpha}a), spatial vegetation patterns do not involve herbivores ($H=0$); they emerge as a stationary periodic pattern ($SP_0$) in a Turing instability of the uniform vegetation state ($UV$) when $P$ decreases below a threshold value $P_T$ (Fig. \ref{fig:bif_diag_alpha}a), as vegetation models without herbivores predict.  
However, at higher $\alpha$ values (Fig.  \ref{fig:bif_diag_alpha}b), the stationary periodic vegetation patterns without herbivores ($SP_0$) lose stability to stationary periodic vegetation patterns with low herbivore density ($SP_H$) as $P$ is increased past a threshold $P_{HS}$. The latter persists as a stable solution in a very small precipitation interval $P_{HS}<P<P_{HT}$, and at $P=P_{HT}$ it loses stability to traveling-wave patterns ($TW_W$). As $P$ is increased, the $TW_W$ solution branch follows very closely the unstable $SP_0$ branch (Fig. \ref{fig:bif_diag_alpha}b).

The mechanism that drives the formation of traveling waves from stationary patterns is illustrated in Fig. \ref{fig:TW_mechanism}. Grazing or browsing at the right edge of a vegetation patch, $x_R$, reduce the vegetation biomass there, $B(x_R)$, and consequently the rate of water uptake. The higher soil-water content at $X_R$ increases soil-water diffusion to the drier patch center. As a result, the left edge of the patch, $X_L$, benefits from reduced water loss by diffusion to the patch center, which favors vegetation growth at that edge.  Thus, while vegetation biomass at the right edge decreases, it increases at the left edge, leading to a traveling vegetation-herbivore wave. These processes are summarized in the loop of negative and positive feedbacks illustrated at the top right part of  Fig. \ref{fig:TW_mechanism}.

At higher precipitation values and higher maximal consumption rates $\alpha$, another solution branch of traveling waves appears, as  Fig. \ref{fig:bif_diag_alpha}c shows. We denote it as $TW_H$, to distinguish it from the traveling-wave solution at low precipitation, denoted by $TW_W$. This traveling-wave solution ($TW_H$) emerges in a non-uniform oscillatory instability of the uniform $UH$ state, even when the water-vegetation scale-dependent feedback is too weak to form patterns, 
indicating it is a result of a scale-dependent feedback associated with herbivores as the dominant inhibitor. 
At these intermediate $\alpha$ values, the $TW_H$ solution is unstable and complex dynamics set in as discussed in Section \ref{sec:complex_dynamics}.

A major difference between the $TW_W$ and $TW_H$ solutions is the much higher herbivore density and spatial spread in the latter. This is because of the wetter conditions that promote vegetation growth and thus support denser herbivore populations. In addition, the traveling-wave speeds of the $TW_H$ solutions are an order of magnitude higher than those of $TW_W$ solutions (typically 0.4m/y vs. 0.06m/y for the parameter values in Table 1), as the sizes of the arrows in Fig. \ref{fig:bif_diag_alpha}e,f indicate. 

\begin{figure}[t]
    \includegraphics[width=0.45\textwidth]{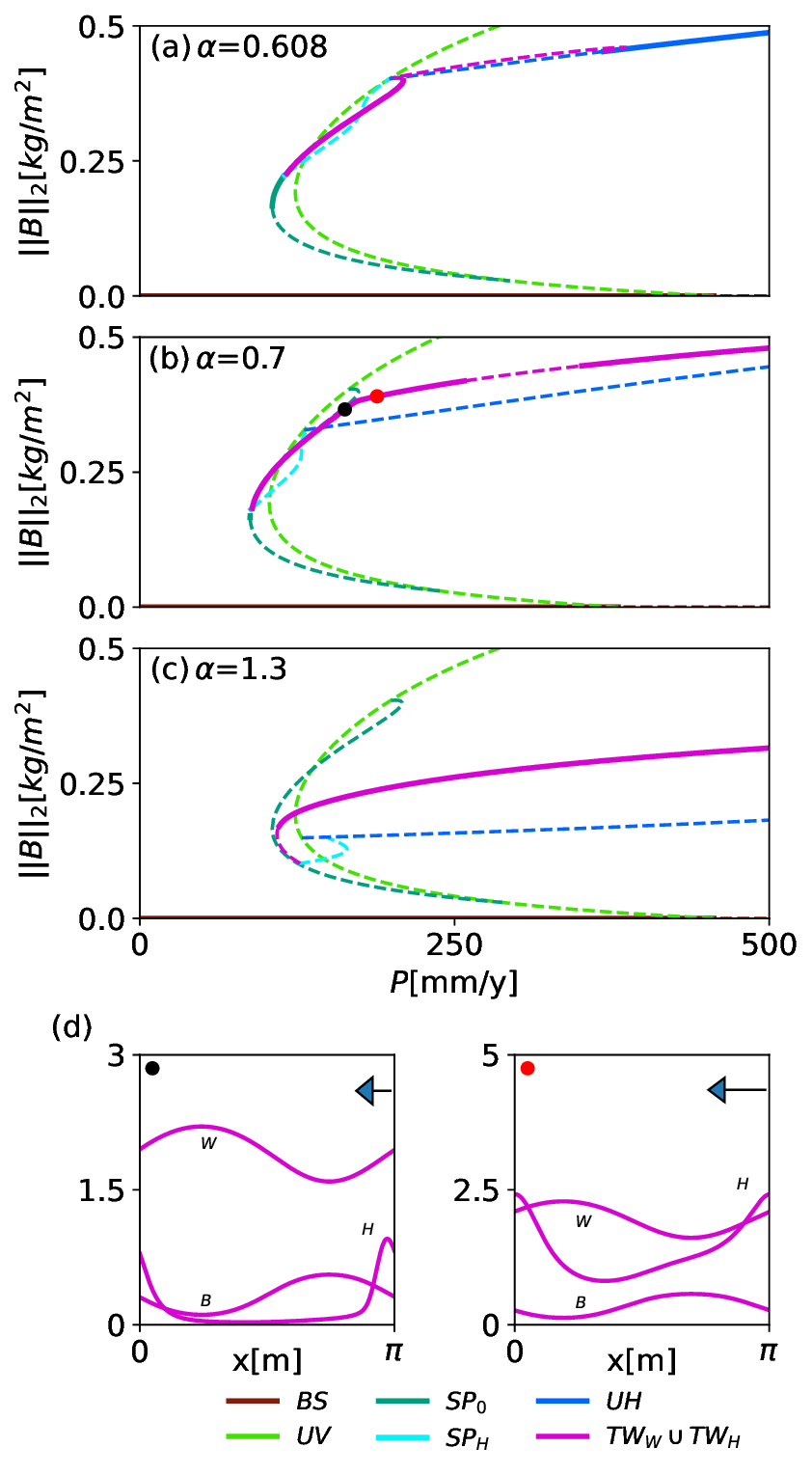}
    \caption{\footnotesize Bifurcation diagrams of uniform and 1d nonuniform solutions of Eq. \eqref{eq:BWH} for intermediate herbivory stress. (a) Merging of the two traveling-wave solutions $TW_W$ and $TW_H$ (see Fig.\ref{fig:bif_diag_alpha}c) into a united solution branch $TW_W\cup TW_H$. (b) At higher $\alpha$ value the transition between the two solutions becomes smooth, yet with a pretty sharp change in the herbivore density and the traveling-wave speed, as the solutions displayed in panels (d) show. The united solution becomes unstable as the precipitation exceeds a threshold value, but that threshold is pushed to higher precipitation values when $\alpha$ is increased, as the bifurcation diagram in (c) shows. Parameters are as indicated and as in Table 1.
    }
    \label{fig:bif_diag_alpha_high}
\end{figure}

According to the bifurcation diagram in Fig. \ref{fig:bif_diag_alpha}a, the herbivore-free uniform vegetation state $UV$ loses stability to the uniform vegetation-herbivore state ($UH$) at $P_H$ and to stationary periodic patterns ($SP_0$) at $P_T<P_H$. As the herbivory stress ($\alpha$) increases, the stability range of $UV$, $P_T<P<P_H$, diminishes and disappears when the two bifurcation points collide ($P_T=P_H$). Beyond that point, that is, at yet higher $\alpha$ values, the two traveling-wave solutions, $TW_W$ and $TW_H$, merge to form a single solution branch,  as Fig. \ref{fig:bif_diag_alpha_high} shows. We denote the merged solution branch $TW_W\cup TW_H$, as $TW$. A closer look at the bifurcation structure across the merging of the two traveling wave solutions is given in Appendix B.
Beyond but close to the merging point of the two traveling-wave solutions, the solutions still retain their identity as panels (b) and (d) in Fig. \ref{fig:bif_diag_alpha} show; localized  herbivore distribution (black circle) vs. wide $TW_H$-like distribution (red circle).

\subsection{Complex traveling-wave dynamics}
\label{sec:complex_dynamics}
\begin{figure}[t!]
    \centering
    \includegraphics[width=0.45\textwidth]{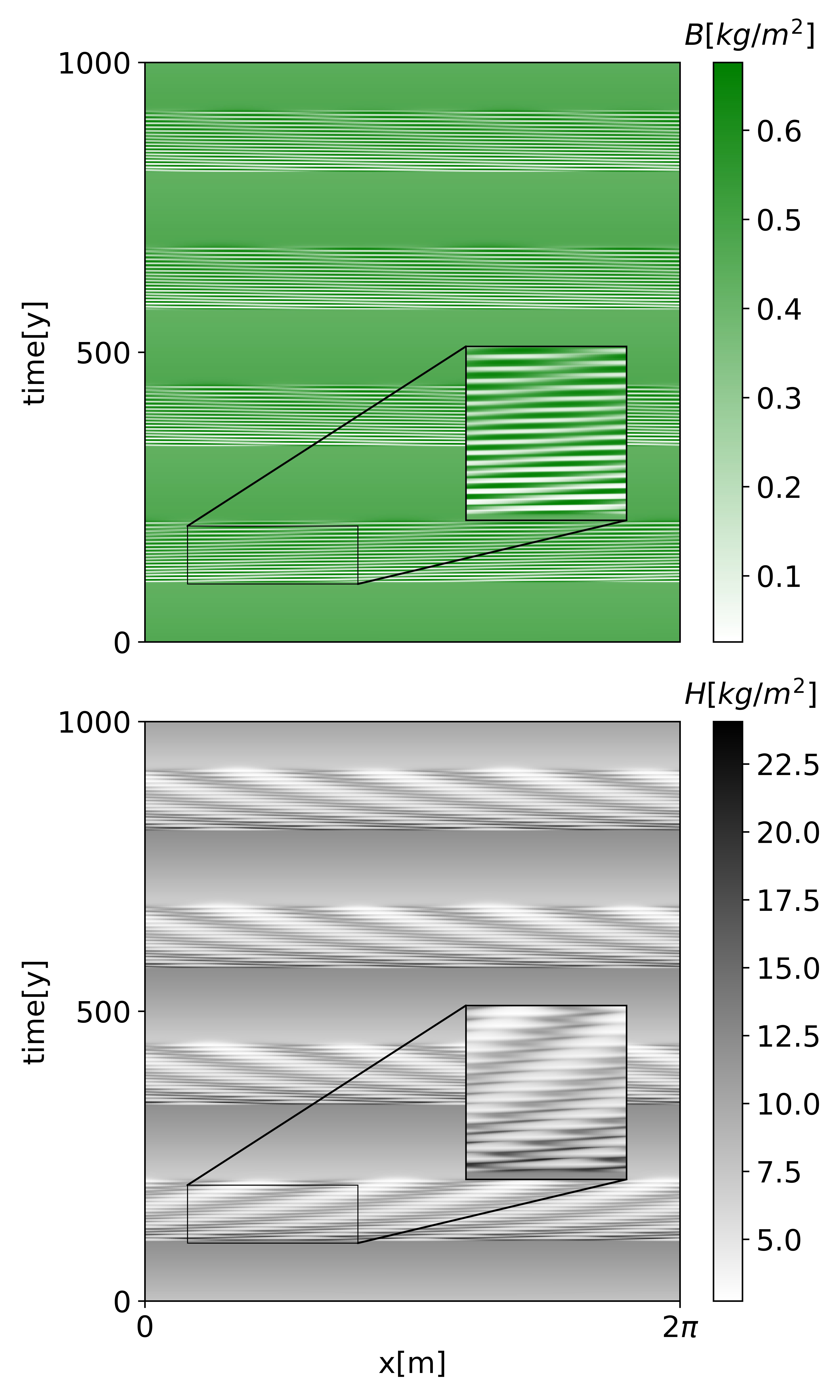}
    \caption{\footnotesize Space-time plots showing complex oscillations at intermediate precipitation range where both the uniform ($UH$) and traveling ($TW_H$) vegetation-herbivore states are unstable. The oscillations involve repeated instances of uniform vegetation $UH$ followed by instances of traveling waves $TW_H$. The color bars on the right denote the values of $B$, $W$ and $H$ in units of kg/m$^2$. Parameter values: $P = 350 mm/y$, $\alpha = 0.608 y^{-1}$ and as in Table 1.}   
    \label{fig:UH_TW_osc}
\end{figure}

The traveling-wave solutions $TW_H$, which bifurcate from the uniform vegetation-herbivore state $UH$, are unstable when they appear, as Fig. \ref{fig:bif_diag_alpha}c shows. This bifurcation results in a precipitation range where both $UH$ and $TW_H$ coexist as unstable solutions. Although the bare-soil solution $BS$ is stable in this range, the system does not necessarily collapse to bare soil but rather exhibits complex oscillations as Fig. \ref{fig:UH_TW_osc} shows. These oscillations involve alternating phases of spatially uniform biomass distributions and traveling waves. During the uniform-distribution phase the herbivore biomass density increases in time. The herbivory stress that builds up induces a transition to traveling waves (similar to the effect of increasing $\alpha$ in Fig. \ref{fig:bif_diag_alpha}). However, the increasing herbivory stress also results in reduced vegetation biomass, which, in turn, leads to a decline in the herbivore biomass density. The consequent reduction in herbivory stress favors a transition back to a uniform distribution and completes an oscillation cycle. The traveling-wave phase involves uneven combinations of left and right traveling waves. 

\section{Patterning and vegetaxis improve herbivores' survival in water-limited systems}
\label{sec:survival}
The sustainability of herbivores in drylands depends to a large extent on the ability of vegetation to tolerate water stress. Water stress increases as the precipitation drops down, but vegetation patterning acts to relax that stress~\cite{Meron2019pt, Rietkerk2021science}, and therefore may contribute to the sustainability of herbivores. The sustainability may also be affected by the foraging strategy of the herbivores, quantified in the model by the strength of vegetaxis. In the following sections we analyze these two factors, vegetation patterning and vegetaxis.

\subsection{Vegetation patterning}
\begin{figure}
    \centering
    \includegraphics[width=0.48\textwidth]{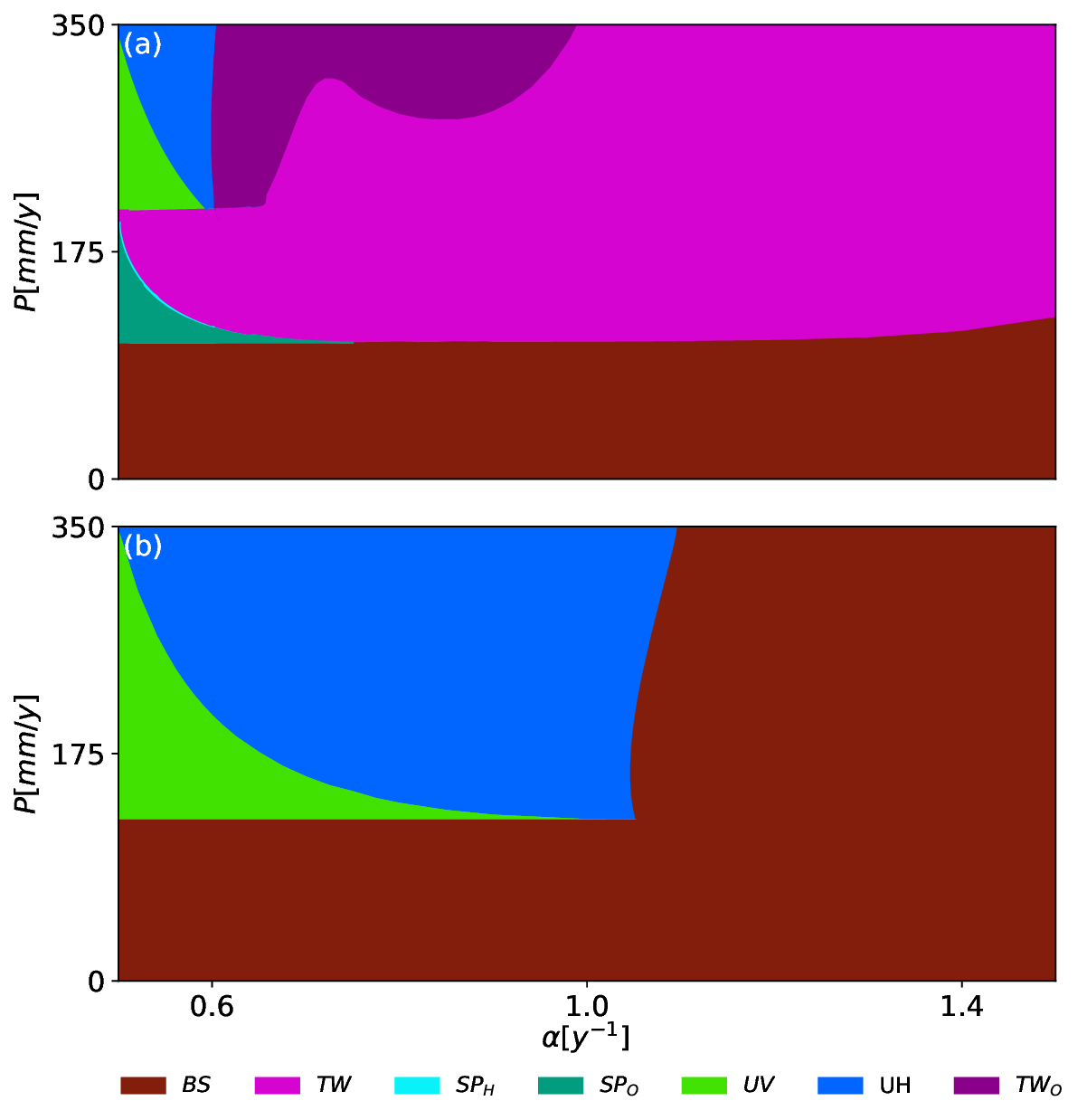}
    \caption{\footnotesize Phase diagrams of stable states in the parameter plane spanned by the precipitation $P$ and maximal consumption rate $\alpha$. Panel (a) shows a phase diagram for the spatial model, where both uniform and nonuniform states are possible. Panel (b) shows, for comparison,  the diagram for a non-spatial model, where nonuniform states are ruled out. Shown are stationary uniform solutions ($UV$, $UH$), stationary periodic patterns ($SP_0$, $SP_H$), periodic traveling waves ($TW$), and traveling-wave oscillations ($TW_O$). The bare-soil state ($BS$) is stable throughout the entire $\alpha$ and $P$ ranges shown, implying wide bistability ranges. Because most of the bifurcations are subcritical, small tristability ranges exist as well. Parameter values are given in Table 1.
    }
    \label{fig:P_alpha}
\end{figure}

The role of vegetation patterning in sustaining herbivores at low precipitation $P$ is clearly evident from the bifurcation diagrams shown in Figs. \ref{fig:bif_diag_alpha} and \ref{fig:bif_diag_alpha_high}: while uniform vegetation can stably sustain herbivores only for $P>P_H$, patterned vegetation can sustain herbivores at significantly lower precipitation values, $P<P_T<P_H$. This conclusion is demonstrated in a more transparent way by the low $\alpha$ range of the phase diagram of stable ecosystem states shown in Fig. \ref{fig:P_alpha}a and its non-spatial counterpart (with no spatial derivative terms) for which patterns are excluded. Traveling vegetation-herbivore waves ($TW$ - magenta domain in  Fig. \ref{fig:P_alpha}a) persist at significantly lower precipitation values than uniform vegetation-herbivore distributions ($UH$ - blue domain in  Fig. \ref{fig:P_alpha}b). 

More striking is the effect of patterning at higher $\alpha$ values; traveling vegetation-herbivore waves (magenta domain) occupy the entire $\alpha$ range considered in Fig. \ref{fig:P_alpha}a, and a wide range beyond it, with the exception of an intermediate behavior (dark-magenta domain) where the system alternates periodically in time between uniform and traveling-wave vegetation-herbivore states (see Fig. \ref{fig:UH_TW_osc}). By contrast, when vegetation patterning is ruled out, collapse to bare soil occurs at relatively low $\alpha$ values, as Fig. \ref{fig:P_alpha}b shows. The positive role traveling waves play in sustaining a functional vegetation-herbivore state at high herbivory stress is a consequence of the uneven stress across each vegetation patch -- high on the trailing side and low on the leading side. The latter is lower than the herbivory stress in the uniform vegetation-herbivore state, and therefore traveling waves sustain higher $\alpha$ values.

We note that the bare-soil state remains stable throughout the whole precipitation range considered in Fig. \ref{fig:P_alpha}, implying wide bistability ranges, as well as  tristability ranges where $UV$ and $TW_W$ are alternative stable states and where $UH$ and $TW_H$ are alternative stable states.
We also note that $\alpha$ values of order unity, as used in Fig. \ref{fig:P_alpha}, represent low consumption rates that can be realized when fodder is supplied as supplementary food. Higher consumption rates describe increasing reliance on herbivory as the primary food resource.


\subsection{Vegetaxis}
\begin{figure}
    \centering
    \includegraphics[width=0.47\textwidth]{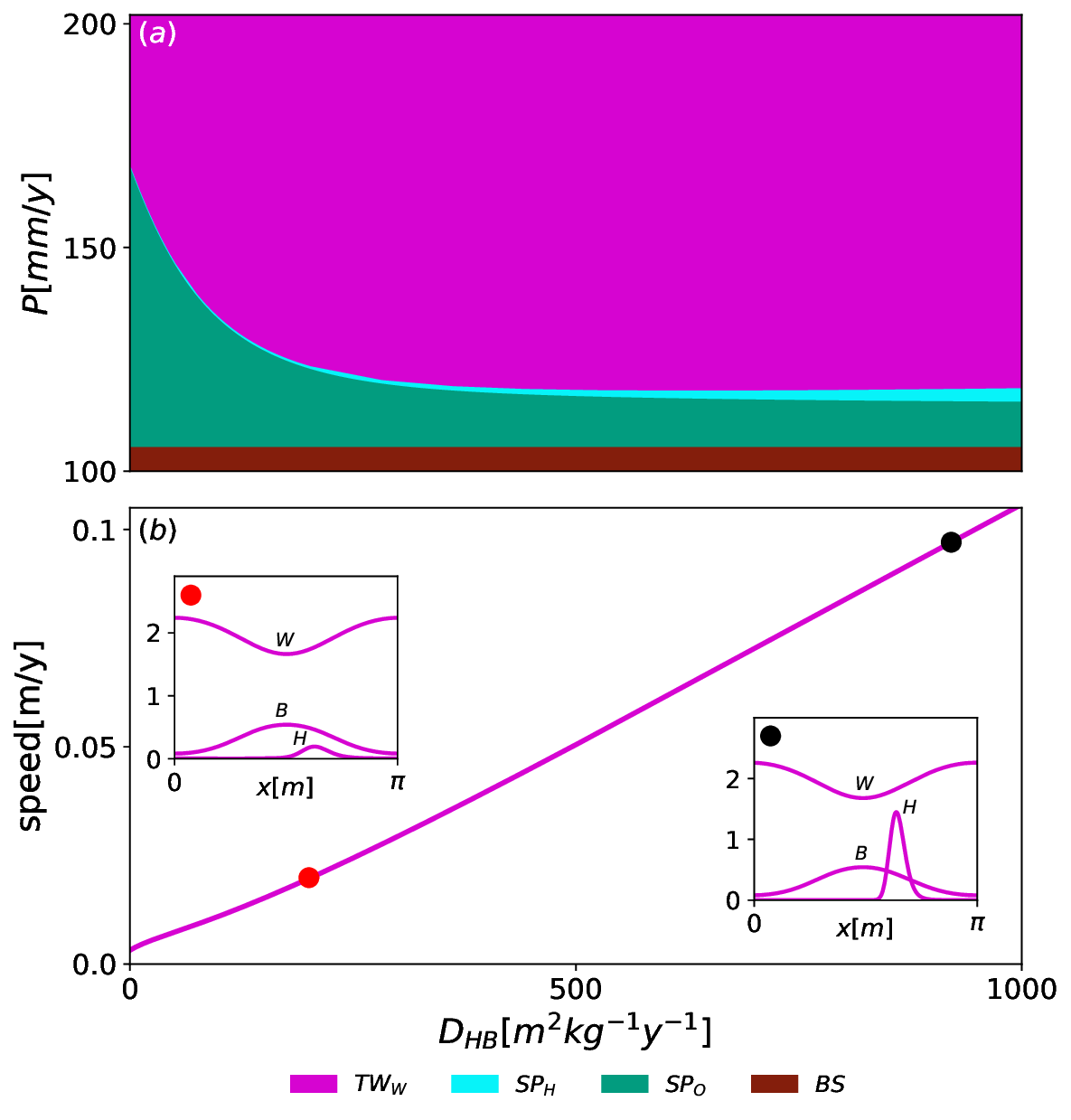}
    \caption{\footnotesize Effects of vegetaxis. (a) A phase diagram of stable ecosystem states in the plane spanned by $D_{HB}$ and $P$. The extended range of $TW_W$ (magenta domain) to lower $P$ values as $D_{HB}$ increases highlights the positive effect that vegetaxis has on herbivore survival. (b) Traveling-wave speed increases as $D_{HB}$ increases. This positive relation is due to the higher herbivore density at larger $D_{HB}$ values, as the two insets show. Parameters:  $P \approx 173 \rm{mm}\cdot \rm{y}^{-1}$ in (b), and as in Table 1.}
    \label{fig:P_DHB}
\end{figure}

The assumed tendency of herbivores to move toward denser vegetation, a behavior we termed `vegetaxis', affects the onset of stationary and traveling patterns involving herbivores. This tendency can be quantified by the parameter $D_{HB}$ that appears in the expression (\ref{eq:D_V}) for the herbivores' motility up vegetation-biomss gradients, $D_V(B)$. Figure \ref{fig:P_DHB}a shows a phase diagram of ecosystem states in the plane spanned by $D_{HB}$ and $P$ in the low precipitation range ($P<250$mm/y). It shows extended stability ranges, toward lower $P$ values, of the herbivore states $TW_W$ and $SP_H$, as $D_{HB}$ increases. This finding highlights the positive role vegetaxis plays in the survival of herbivores under conditions of water stress. 

The positive role of vegetaxis in tolerating water stress can be understood in the following way; in the absence of vegetaxis ($D_{HB}=0$), herbivores perform a random walk, accounted for by the first term in the flux (\ref{eq:J_H}), and may miss nearby vegetation patches. The longer time they spend in bare soil, lacking food, increases their mortality. By contrast, in the presence of vegetaxis, herbivores sense the tails of nearby vegetation patches, slow down and move toward the patches to graze. This behavior represents an exploitation strategy~\cite{Berger-Tal2014plosone}, where the herbivores exploit their knowledge that small vegetation-biomass values are indicative of nearby vegetation patches and slow down to graze even though there might be bigger and denser vegetation patches farther away.


Another outcome of vegetaxis is an increased traveling-wave speed, as Fig. \ref{fig:P_DHB}b,c shows. As $D_{HB}$ increases more herbivores accumulate at the edge of a vegetation patch. The resulting higher herbivore intensity and its positive effect on the soil-water content by reduced uptake act to increase the traveling-wave speed. 
The increased traveling-wave speed may have significant implications for biodiversity, as discussed in Section \ref{sec:Discussion}.

\subsection{Collective herbivore dynamics}
\label{sec:collective}
The time-dependent spatial distribution of the herbivores, represented by $H(X,Y,T)$, provides information about collective herbivore dynamics in a mean-field sense. 
The major processes that shape these dynamics in the model are herbivore diffusion (random walk) and vegetaxis with biomass-dependent motilities, as the expression  (\ref{eq:J_H}) for the herbivore flux $J_H$ shows, and the increase of herbivory stress as herbivores reproduce and their density increases.


\begin{figure}
    \includegraphics[width = 0.5\textwidth]{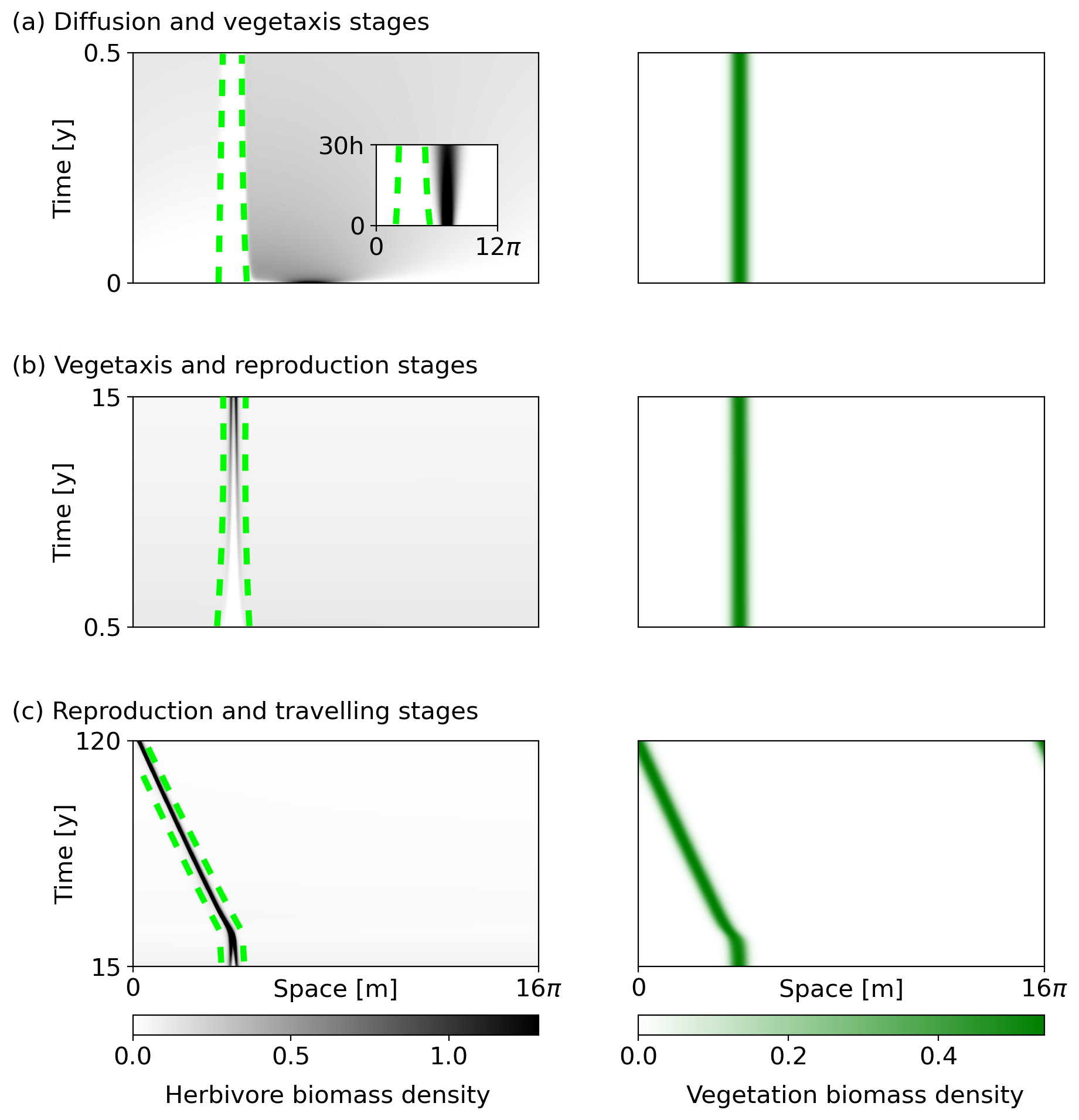}
    \caption{\footnotesize 
   Space-time plots of herbivore biomass (left) and vegetation biomass (right), depicting the different processes that shape their dynamics. The green dashed lines on the left show the boundaries of the vegetation patch. (a) Starting with localized herbivore distributions away from vegetation patches (periodic boundary conditions are used), symmetric herbivore diffusion, representing random walk, takes place as the inset shows. At later times vegetaxis begins. (b) Vegetaxis leads to the accumulation of herbivores at the vegetation-patch boundaries, but more on the (right) boundary that is closer to the initial herbivore distribution. In the course of time, reproduction increases the herbivore density at the boundaries. (c) The increased herbivore density by reproduction induces a transition to a traveling wave. Parameters: 
   $P$=110 mm/y, and as in Table 1.
    }
    \label{fig:TW_spacetime}
\end{figure}

We consider first the dynamics of herbivore aggregation at vegetation patches for low herbivory stress for which the asymptotic herbivore distribution is localized at one side of the patch (see Fig. \ref{fig:bif_diag_alpha}e). We start with
a single vegetation patch and localized herbivore distribution sufficiently far from the patch, not to ``sense'' its existence as Fig. \ref{fig:TW_spacetime}a shows. Four stages can be distinguished in the subsequent dynamics: (i) symmetric herbivore spread by random walk (diffusion), (ii) vegetaxis by a low-density population of herbivores who sense the vegetation patch, (iii) further herbivore accumulation and growth at the vegetation patch, (iv) vegetation-patch displacement (Fig. \ref{fig:TW_mechanism}) as the herbivores' density increases, leading to a constant-speed traveling wave. These stages and the transitions from one stage to another are demonstrated in Fig. \ref{fig:TW_spacetime}.

\begin{figure*}
    \includegraphics[width = 1\linewidth]{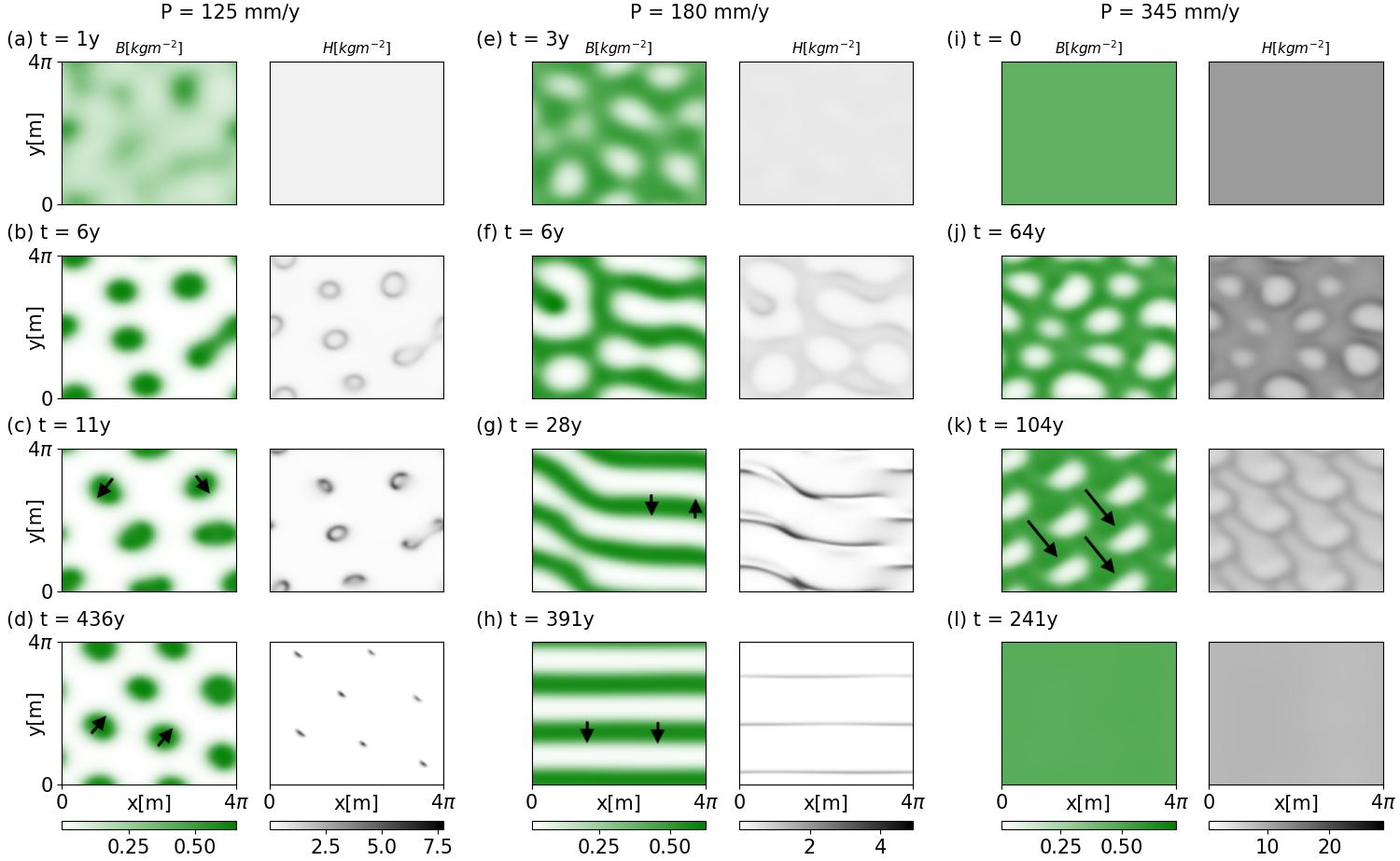}
    \caption{\footnotesize 
   Formation of traveling-wave patterns for low herbivory stress (low $\alpha$). In all simulations the initial state is randomly perturbed uniform vegetation and herbivore distributions. (a-d) Formation of traveling spot pattern at $P=125 mm/y$. (e-h) Formation of traveling stripe patterns at $P=180 mm/y$. In both cases, initially, a nearly stationary vegetation pattern forms while the herbivore distribution is still fairly uniform (a,e). Then, herbivores in sparse vegetation areas move toward the denser vegetation patches (spots or stripes), accumulating at their edges (b,f). At later times, the herbivores concentrate at one edge, forming patches (spots, stripe segments) that travel in different directions, indicated by the arrows (c,g). At yet later times a traveling-wave pattern forms moving as a whole in one direction (d,h). (i-l) Development of disordered dynamics involving oscillations between uniform vegetation-herbivore distributions and traveling gap patterns at $P=345 mm/y$. Unlike the dynamics at lower precipitation values (a-h), vegetation and herbivore patterns emerge simultaneously from uniform distributions (j), and the herbivore distributions are spatially extended rather than localized (i-l). See text for further explanations. Time values are given in years. Parameter values are as in Table 1. 
    }
    \label{fig:2d_simulations}
\end{figure*}

In two space dimensions (2d), the 1d traveling-wave solutions discussed so far represent constant-speed traveling stripe patterns. These patterns appear at intermediate precipitation values, as Fig. \ref{fig:2d_simulations} shows. At lower precipitation, traveling spot patterns emerge, while at higher precipitation, oscillations between uniform vegetation-herbivore distributions and traveling gap patterns emerge, corresponding to the $TW_O$ solution in 1d. At yet higher precipitation traveling gap patterns appear (not shown in Fig. \ref{fig:2d_simulations} but similar to the traveling gap pattern shown in Fig. \ref{fig:2d_simulations}k). In all simulations shown, the initial conditions are  uniform vegetation-herbivore distributions subject to random spatial perturbations. 

\

The simulations reveal several aspects of vegetation-herbivore dynamics. At low precipitation ($P=120$ mm/y and 180 mm/y), vegetation patterns appear before herbivore patterning, while at high precipitation ($P=345$ mm/y), these patterns appear simultaneously. Furthermore, at low precipitation, the asymptotic herbivore distributions are localized at patch edges with very sparse distribution elsewhere, while at high precipitation the herbivore distribution is more spatially extended. These behaviors are consistent with the 1d continuation results shown in Fig. \ref{fig:bif_diag_alpha_high}d. They reflect the different pattern formation mechanisms that prevail at low and high precipitation; patterning driven by water stress at low precipitation and by herbivory stress at high precipitation, as the denser vegetation at high precipitation supports denser herbivore populations.  

\begin{figure*}
    \includegraphics[width = 0.8\linewidth]{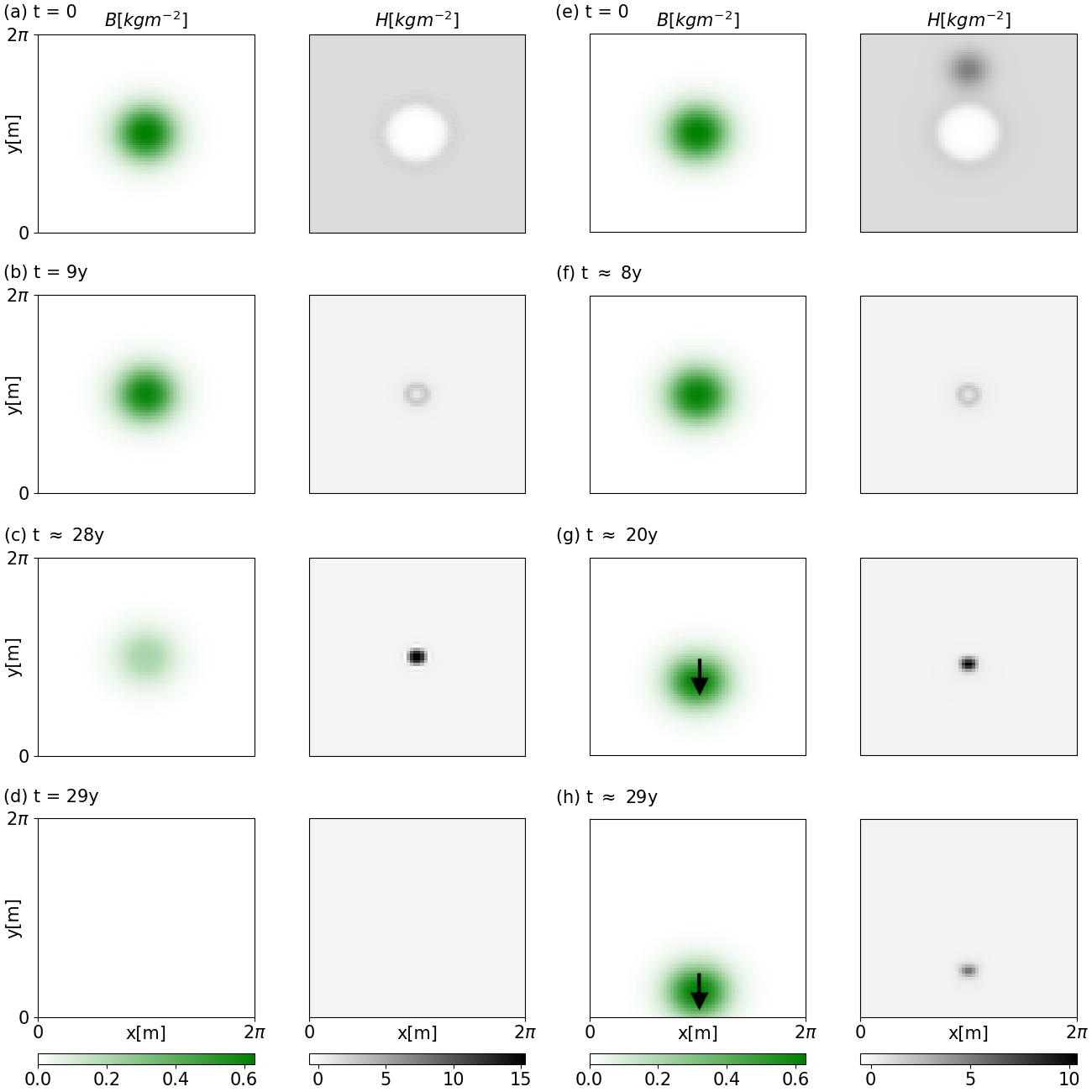}
    \caption{\footnotesize 
   Vegetation-patch survival by asymmetric grazing that leads to a traveling patch. (a-d) A symmetric herbivore distribution (gray) around a vegetation patch (green) results in complete consumption and collapse of both plant and herbivore populations. (e-h) Breaking the circular symmetry by increasing the herbivore density on one side of the vegetation patch results in a vegetation patch of fixed size traveling at a constant speed, as the arrow indicates. Time values are given in years. Parameter values are as in Table 1. 
   }
    \label{fig:single_patch}
\end{figure*}

The simulations shown in Fig. \ref{fig:2d_simulations} also reveal two interesting aspects of herbivore dynamics. At first, herbivores appear to approach the entire edge of each vegetation patch (Fig. \ref{fig:2d_simulations}b,f,j) but then concentrate on one side, forming a traveling patch (Fig. \ref{fig:2d_simulations}c,g,k). These behaviors lead to disordered dynamics as different vegetation patches generally move in different directions, but eventually, the dynamics are synchronized, forming patterns that travel as a whole in one direction (Fig. \ref{fig:2d_simulations}d,h,k). The symmetry breaking associated with herbivore concentration on one side of each vegetation patch is further investigated in Fig. \ref{fig:single_patch}, where a comparison between the dynamics of a perturbed and unperturbed circular vegetation patch is made. While symmetric herbivory at the edges of an unperturbed circular patch results in collapse to bare soil, asymmetric herbivory in the perturbed patch results in patch survival in the form of a traveling vegetation-herbivore patch.

\section{\label{sec:Discussion}Discussion}
We introduced here a new model that allows studying the sustainability of dryland pastures under conditions of increased water and herbivory stress. Water stress increases as drier climates develop, and herbivory stress increases as the demand for food production due to population growth extends. The formation of stationary vegetation patterns is crucial in sustaining water stress. Using the model, we find that the formation of traveling waves plays a similar role in sustaining combined water-herbivory stress. 

Two types of traveling-wave solutions are identified: (i) Nearly stationary slow traveling solutions, appearing in an oscillatory instability of stationary vegetation patterns at low precipitation rates ($TW_W$) and describing small-amplitude localized herbivore distributions. (ii) Fast traveling solutions, appearing in an oscillatory instability of stationary uniform vegetation at higher precipitation rates ($TW_H$) and describing large amplitude spatially extended herbivore distributions. As the grazing stress is increased, the two solution branches merge into a single branch along which the traveling-wave speed and the herbivore distribution sharply change but in a continuous manner. 
The traveling-wave nature of vegetation-herbivore patterns may have significant implications for biodiversity, ecosystem function, and ecosystem management, as discussed below. 

Traveling vegetation-herbivore waves may affect species diversity in two ways. Unlike stationary patterns that provide ample time for high-fitness plant species to out-compete lower-fitness species, traveling waves provide a limited period of competition time at any given location and, therefore, are expected to support higher species diversity. 
Another way by which traveling vegetation-herbivore waves may affect species diversity is through ameliorated growth conditions associated with plant mortality and the formation of temporal bare-soil patches. Increased dead organic matter and soil biota~\cite{Moore2004eco_let} in these patches may favor the growth of faster-growing species that would not grow in eroded permanent bare-soil patches associated with stationary patterns~\cite{Eldridge2011eco_let}.


A variety of stable uniform states and periodic patterns, stationary and traveling, exist in precipitation ranges where bare soil is also a stable state ($P<P_B$), as the bifurcation diagrams in Fig. \ref{fig:bif_diag_alpha} and the phase diagram in Fig. \ref{fig:P_alpha} show. In these ranges, early tipping to the dysfunctional bare-soil state is possible, despite the co-existence of stable functional vegetation-herbivores states. A possible scenario of early tipping is a severe and prolonged drought developing over a relatively short time that pushes the system out of the attraction basin of its state. Interfering with this so-called rate-dependent tipping or R-tipping~\cite{Feudel2018chaos,Ritchie2023esd} are invariant manifolds associated with unstable states that may constrain the response and prevent tipping~\cite{Zelnik2021plos_comp_biol}. A deeper understanding of these dynamics is needed for devising effective intervention forms to evade tipping, such as managing grazing non-uniformly in space, or grazing-stress control~\cite{Vignal2023eco_evo}. Figure \ref{fig:single_patch} hints at an intervention form that favors convergence to a functional vegetation-herbivore traveling state over collapse to bare soil, namely, managing herbivory in a way that breaks the symmetry along the edges of vegetation patches. 


The model presented here provides a platform of model variants that can be used to study a variety of additional questions. Vegetaxis, as modeled here, represents an exploitation strategy, where herbivores use the information they acquire by sensing vegetation-density gradients to identify nearby vegetation patches and graze or browse there, even though denser patches may exist farther away. A simple model variant can also capture elements of exploration strategy, whereby herbivores do not necessarily graze or browse at their first encounter with a vegetation patch but rather explore the area for denser vegetation patches. Modeling this combined exploration-exploitation strategy may require modification of the vegetaxis term $D_V(B)$ (Eq. (\ref{eq:D_V})) or its parameters' values, and the introduction of soil heterogeneity, e.g., to model soil-rock mosaics~\cite{Sheffer2013eco_let}, that results in vegetation patches of different densities. A question of interest here is how different exploration-exploitation strategies affect the sustainability of the system under combined water-herbivory stress.
Another question that can be studied is the effect of traveling vegetation-herbivore waves on the diversity of the plant community. To this end, the model can be extended to include a trait subspace representing the pool of functional traits that characterize the community~\cite{Bera2021elife}. These and other examples point toward the model's utility to study various new questions associated with dryland pastures at risk of desertification.

\section*{Acknowledgement}
We thank Shachar Feder for helpful discussions. This research has received funding from the Israel Science Foundation under grant No. 2167/21, and from the European Union’s Horizon Europe – European Research Council programme under ERC-2022-SYG Grant Agreement No 101071417 - RESILIENCE.

\appendix
\section{Non-dimensional form of the model equations}
It is instructive to transform the model equations (\ref{eq:BWH}) to a non-dimensional form in order to uncover equivalence aspects of different parameters. The model represents an $\cal{LMT}$ system, where $\cal{L}$, $\cal{M}$, and $\cal{T}$ stand for the dimensions of length, mass, and time, respectively~\cite{Meron2015book}. To rescale all model quantities to non-dimensional forms, we need to choose three dimensionally-independent parameters. The choice we make is $M_H$, $\Lambda$, and $D_B$, whose independent dimensions are ${\cal{T}}^{-1}$, ${\cal{L}}^2 {\cal M}^{-1}{\cal T}^{-1}$, and ${\cal{L}}^2 {\cal T}^{-1}$, respectively. In terms of these dimensions, the dimensions of all other quantities in the model can be expressed as described below. We choose to denote the non-dimensional forms of the independent variables ($X,Y,T$) and the dependent variables ($B,W,H$) by the corresponding lower-case letters, and the non-dimensional forms of the parameters by adding a tilde sign.

The non-dimensional model then reads:
\begin{align}
    \begin{split}
        \partial_tb&= bw(1+\tilde{E}b)^2(1-b/\tilde{K}_B)-\tilde{M}_B b-\frac{\tilde{\alpha}bh}{\tilde{\beta}+b} +\nabla^2b\,,\\
        \partial_tw&=\tilde{P}-\frac{\tilde{N}w}{1+Rb/\tilde{K}_B}- \tilde{\Gamma} bw(1+\tilde{E}b)^2 + \tilde{D}_W\nabla^2w\,,\\
        \partial_th&=-h+ A\frac{\tilde{\alpha}bh}{\tilde{\beta}+b}{\left(1-h/\tilde{K}_H\right)} -\nabla\cdot \tilde{J}_H\,,\\
    \end{split}
    \label{eq:BWH_app}
\end{align}
where
\begin{equation*}
\label{eq:J_H_app}
    \tilde{J}_H = -\tilde{D}_{HH}\frac{\tilde{\xi}^2}{\tilde{\xi}^2+b^2}\nabla h + h\tilde{D}_{HB}\frac{\tilde{\kappa}}{\tilde{\kappa}+b}\nabla b\,, \\
\end{equation*}
with $\nabla=\hat{\textbf{x}}\partial_x+\hat{\textbf{y}}\partial_y$, where $\hat{\textbf{x}}, \hat{\textbf{y}}$ are unit vectors in the $x,y$ directions.
The non-dimensional independent variables are given by 
\begin{equation*}
 t=M_H T,\quad x=\sqrt{M_H/D_B}X,\quad y=\sqrt{M_H/D_B}Y\,,
\end{equation*}
and the non-dimensional dependent variables by 
\begin{equation*}
    b=\Lambda B/M_H,~ w=\Lambda W/M_H,~ h=\Lambda H/M_H\,.
\end{equation*}

\begin{figure}[t]
    \includegraphics[width=0.45\textwidth]{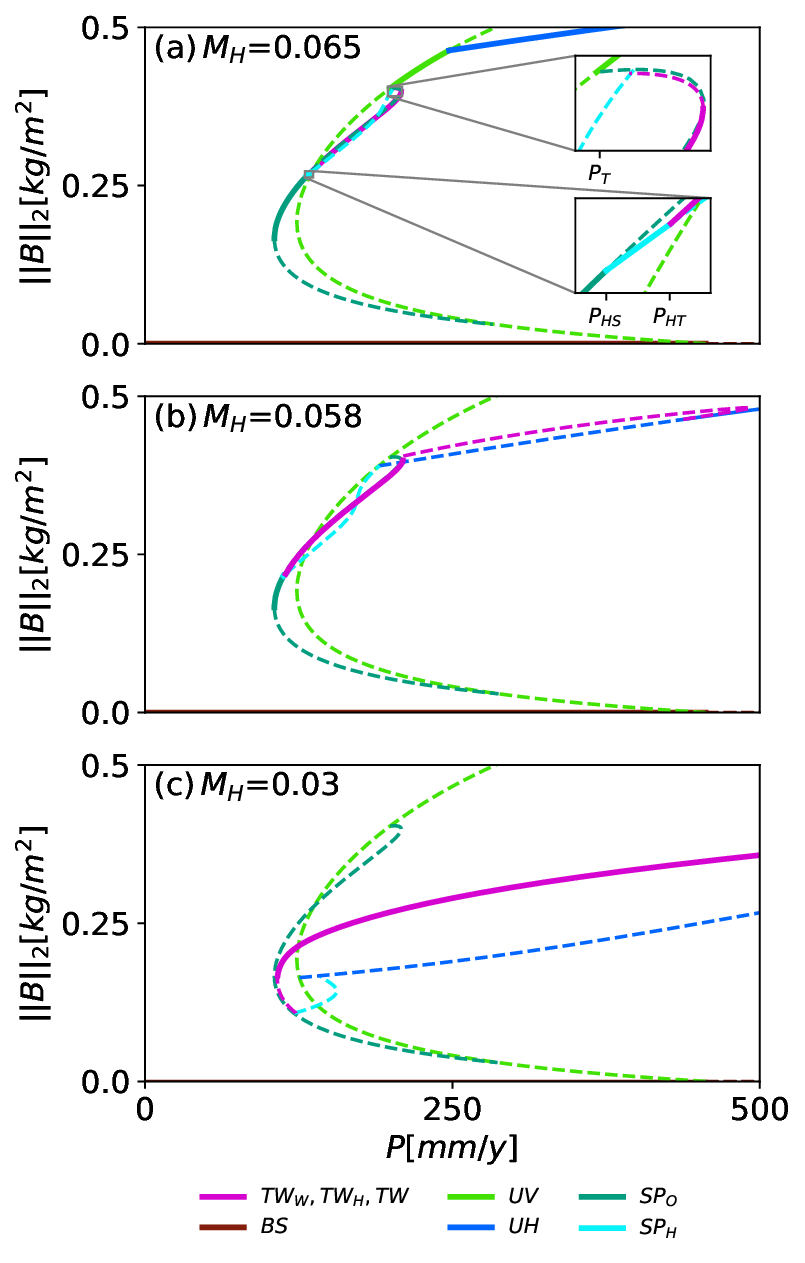}
    \caption{\footnotesize 
   Herbivore mortality has the opposite effect on the bifurcation structure that herbivory stress has. Shown are bifurcation diagrams of uniform and 1d nonuniform solutions of Eq. \eqref{eq:BWH} at decreasing $M_H$ values. The two separate traveling solution branches, $TW_W$ and $TW_H$, at high $M_H$ values merge to form a single solution branch, $TW=TW_W\cup TW_H$, at sufficiently low $M_H$ values, similarly to the solutions merge as $\alpha$ is increased (see Figs. \ref{fig:bif_diag_alpha} and \ref{fig:bif_diag_alpha_high} and Appendix B). 
    }
    \label{fig:bif_diag_M_H}
\end{figure}

\noindent The non-dimensional parameters are given by:
\begin{eqnarray*}
\tilde{E}&=&M_HE/\Lambda, ~\tilde{K}_B=\Lambda K_B/M_H, ~
\tilde{M}_B=M_B/M_H\,,\\
\tilde{\alpha}&=&\alpha/M_H,~ \tilde{\beta}=\Lambda\beta/M_H,~ 
\tilde{P}=\Lambda P/M_H^2\,,\\
\tilde{N}&=&N/M_H,~  \tilde{\Gamma}=\Gamma/\Lambda,~ \tilde{D_W}=D_W/D_B\,,\\
\tilde{K}_H&=&\Lambda K_H/M_H,~ \tilde{D}_{HH}=D_{HH}/D_B,~ \tilde{\xi}=\Lambda\xi/M_H\,,\\
\tilde{D}_{HB}&=&M_HD_{HB}/(\Lambda D_B),~ \tilde{\kappa}=\Lambda\kappa/M_H.
\end{eqnarray*}
The non-dimensional forms of the model parameters suggest possible similarities in the effects of different parameters on the bifurcation structure and dynamic behaviors. For example, the form of $\tilde{\alpha}$ suggests that decreasing $M_H$ may have a similar effect to increasing $\alpha$. The bifurcation diagrams in Fig. \ref{fig:bif_diag_M_H}, obtained by decreasing $M_H$, indeed show the same structures and structure changes as the bifurcation diagrams shown in Figs. \ref{fig:bif_diag_alpha} and \ref{fig:bif_diag_alpha_high}, which were obtained by increasing $\alpha$.

Similarly, the non-dimensional form, $\tilde{\beta}$, suggests that increasing the reference vegetation biomass for herbivore satiation, $\beta$, has a similar effect to decreasing $M_H$ or increasing $\alpha$. 


\begin{figure}[t!]
    \centering
    \includegraphics[trim=0.3cm 0cm 0cm 0cm, width=0.49\textwidth]{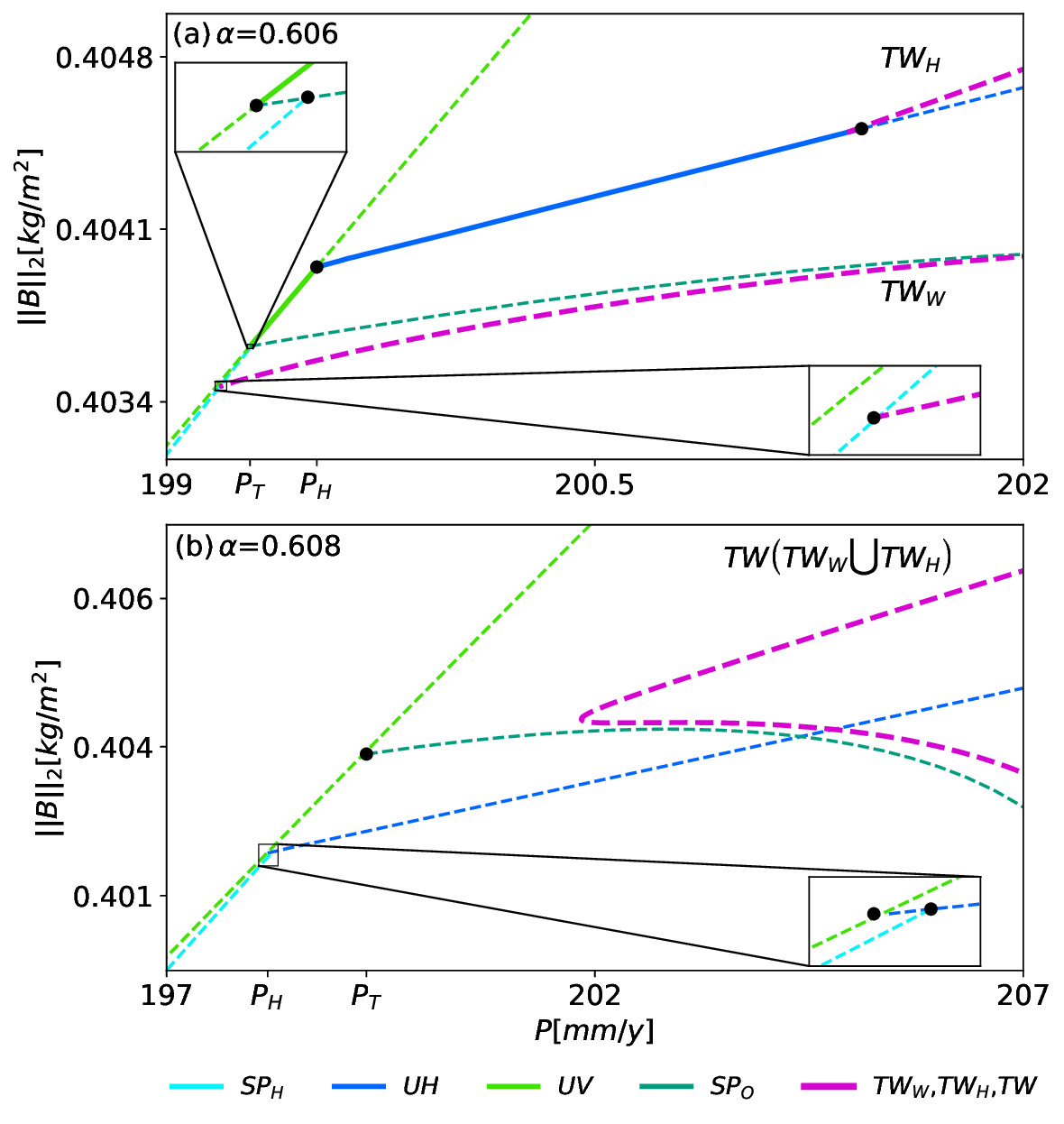}
    \caption{\footnotesize A closer look at the merging of the separate traveling solution branches, $TW_W$ and $TW_H$, at low herbivory stress into a single solution branch, $TW=TW_W\cup TW_H$, at high herbivory stress. The merge occurs as the bifurcation points $P_T$ and $P_H$ exchange locations. See the text in Appendix B for further details. 
    }
    \label{fig:TW_merging}
\end{figure}

\section{The merging of traveling-wave solutions}
Two traveling-wave solutions are distinguished at low herbivory stress (low $\alpha$ values): $TW_W$ at low precipitation rates, where water scarcity is the dominant inhibitor, and $TW_H$ at higher precipitation rates, where herbivores are the dominant inhibitor. As $\alpha$ is increased, the two solution branches merge together to form a single branch ($TW$). This behavior appears to happen when the uniform instability of the herbivore-free uniform-vegetation state $UV$ to the uniform vegetation-herbivore state $UH$ coincides with the Turing instability of $UV$ to the herbivore-free stationary periodic pattern state $SP_0$; that is, at $P=P_H=P_T$ (see Fig. \ref{fig:bif_diag_alpha}). Figure \ref{fig:TW_merging} shows a closeup of the bifurcation structure slightly below the merging point (a), at $\alpha=0.606$, 
and slightly above it (b), at $\alpha=0.608$. 
Below the merging point $P_T<P_H$, and both uniform states, $UV$ and $UH$, have stability ranges (solid green and blue lines). Above the merging point $P_T>P_H$, and no stability ranges of $UV$ and $UH$ exist. Furthermore, as the merging point is traversed, the stationary vegetation-herbivore pattern solution $SP_H$ reconnects; for  $P_T<P_H$ (Fig. \ref{fig:TW_merging}a), it bifurcates from the stationary herbivore-free periodic solution $SP_0$, whereas for $P_T>P_H$, it bifurcates from the uniform vegetation-herbivore solution $UH$ (Fig. \ref{fig:TW_merging}b). 



\bibliography{BWH.bib}
\end{document}